\documentstyle[aps]{revtex}

\begin{document}
\draft
\twocolumn[\hsize\textwidth\columnwidth\hsize\csname
@twocolumnfalse\endcsname
\date{May 16, 1999}
\title{Thermodynamic Geometry and \\ Locally Anisotropic
 Black Holes}
\author{Sergiu I. Vacaru}
\address{Institute of Applied Physics, Academy of Sciences,\\
5 Academy str., Chi\c sin\v au\ MD2028, Republic of Moldova \\ {--- } \\
 Electronic address: vacaru@lises.asm.md  }
\maketitle
\vskip0.2cm
\begin{abstract}
Thermodynamic properties of locally anisotropic $(2+1)$--black holes
are studied by applying geometric methods. We consider a new class
of black holes with a constant in time elliptical event horizon which
is imbedded in a generalized Finsler like spacetime geometry induced
from Einstein gravity. The corresponding thermodynamic systems are
three dimensional with entropy $S$ being a hypersurface function on
mass $M,$ anisotropy angle $\theta$  and eccentricity of elliptic
deformations $\varepsilon .$ Two--dimensional curved thermodynamic
geometries for locally anistropic deformed black holes are constructed
after integration on anisotropic parameter $\theta$. Two approaches,
the first one  based on two--dimensional hypersurface parametric
geometry and the second one developed in a Ruppeiner--Mrugala--Janyszek
fashion, are analyzed.  The thermodynamic  curvatures are computed and
the critical points of curvature vanishing are defined.
 \end{abstract}

\vspace{0.2cm}
 ]

\section{Introduction}

This is the second paper in a series in which we examine black holes for
spacetimes with generic local anisotropy (la). In the first paper (hereafter
referred to as paper I) we analyzed the low--dimensional locally anisotropic
gravity (in brief, we shall use terms like la--gravity, la--spacetime,
la--geometry, black la--holes and so on) and constructed new classes of
locally anisotropic $2+1$--black hole solutions \cite{v4}.

In particular, it was shown following \cite{v1,v2,v3} how black holes can
recast in a new fashion in generalized Finsler--Kaluza--Klein spaces and
 emphasized that such type solutions can be considered in the
 framework of usual
Einstein gravity on anholonomic manifolds. We discussed the physical
properties of $(2+1)$--dimensional black holes with la--matter, induced by a
rotating null fluid and by an inhomogeneous and non--static collapsing null
fluid, and examined the vacuum polarization of la--spacetime by
non--rotating black holes with ellipsoidal horizon and by rotating black
la--holes with time oscillation and ellipsoidal horizons. We concluded that
a general approach to the black la--holes should be based on a kind of
nonequilibrium thermodynamics of such objects imbedded into la--spacetime
ether being a continuous with possible dislocations and declinations.
Nevertheless, we proved that for the simplest type of black la--holes theirs
thermodynamics could be defined in the neighborhoods of some equilibrium
states when the horizons are deformed but constant with respect to a frame
base locally adapted to the nonlinear connection structure which model the
local anisotropy.

In this paper we will specialize to the geometric thermodynamics of, for
simplicity non--rotating, black la--holes with elliptical horizons. We
follow the notations and results from paper I (see also the Appendix) which
are reestablished in a manner compatible in the locally isotropic
thermodynamic \cite{cai} and spacetime \cite{btz} limits with the
Banados--Teitelboim--Zanelli (BTZ) black hole. We defer the examination of
higher dimension and string locally anisotropic black holes to a third paper
(III) \cite{v3}. Here we also remark that a paper under preparation (IV)
 \cite{v5} is devoted to the nonequilibrium thermodynamics of black
 la--holes. This new approach (to black hole physics) is possible
 for la--spacetimes and is based on classical results
\cite{gp,klim,kre,sien}.

We emphasize a postulate which is assumed (in non--explicit form) in general
relativity:\ {\bf the locally an\-isot\-ropic
 distributed matter gives rise to a
locally isotropic geometry.} This is contained in the structure of Einstein
equations for a metric $g_{ij}(x^{k}) $ on a (pseudo) Riemannian space:
\begin{eqnarray}
& G_{ij}(x^{k}) \simeq & \nonumber \\
& \fbox{\sf \ the Einstein tensor for a locally isotropic spacetime \ }
 & \nonumber \\
 & \Updownarrow & \nonumber \\
 & T_{ij}(x^{k},y^{a}),&  \nonumber \\
& \fbox{\sf the energy--momentum tensor of anisotropic matter}
 \nonumber
\end{eqnarray}
where $x^i, i=1,...,n$ are coordinates on spacetime $M$ and $y^a,
a=1,2,...,m,$ are parameters (coordinates) of anisotropies.
So, the Einstein theory formulated in the framework of
(pseudo) Riemannian geometry  has been considered to be locally
 isotropic. Usually, anisotropies are considered as induced
 by matter distributions, quantum fluctuations during inflation and
so on. It is obvious that an energy--momentum tensor for a locally
anisotropic  matter (depending on some anisotropy parameters)
being proportional to the Einstein tensor must induce corresponding
anisotropies of the metric if the theory is required to be self--consistent.
This topic of construction of solutions of systems of Einstein--matter
 fields equations with both dependencies on anisotropic parameters
 of metric and matter fields is usually omitted in general relativity.

Metrics with local anisotropy (depending on tangent vectors) of type%
$$
g_{ij}=\frac 12\frac{\partial ^2F^2\left( x^l,y^k\right) }{\partial
y^i\partial y^j}
$$
where
$$
F\left( x^l,\lambda y^k\right) =\lambda F\left( x^l,y^k\right) ,
$$
$\lambda $ is a real number, i.e the function $F(x^l,y^k)$ is homogeneous on
variables $y^k,$ were fist considered by Finsler and usually they are
associated to the so--called Finsler geometry provided with
 nonlinear and linear connections and metrics structures and respectively
 computed various types of curvatures and torsions (see references \cite{fin}
and on further geometric generalizations and developments with applications
in physics one could consult \cite{ma,v1,v2,v3,vg}). It was considered that
such type geometries are less suitable for direct applications in physics
because of substantial problems with definition of conservation laws
(without local symmetries it was not clear in which manner one could define,
for instance, energy momentum type values), absence of well backgrounded
physical arguments and complexity of such type geometries.

Nevertheless, the very sceptic attitude was changed after it was proved that
Finsler like geometries and their generalizations (the so called generalized
Lagrange spaces \cite{ma}), as well physical theories with generic locally
anisotropic interactions and/or Kaluza--Klein models, can be modelled in a
unified geometrical manner on vector bundles provided with nonlinear
connection structures (a subclass of anholonomic manifolds; as a general
introduction into the spacetime differential geometry we cite \cite{haw} and
on the geometry of nonlinear connections one could consult references \cite
{ma} and \cite{v1,v2}).

The field equations of locally anisotropic gravity are of type
$$
\fbox{%
$
G_{\alpha \beta }(x^i ,y^a ) \simeq T_{\alpha \beta }(x^i , y^a ) $}
$$
where the Einstein tensor is defined on a bundle (generalized Finsler)
space, $x^i $ are usual coordinates on the base manifold and $y^b $ are
coordinates on the fibers (parameters of anisotropy). We emphasize that in
general base and fiber spaces are of different dimensions, i.e. $\dim
\{x^i\}\neq \dim \{y^b \}.$ As a matter of principle this is a usual
Einstein theory but on generic anholonomic manifolds (vector bundles) with
 a locally adapted nonlinear connection (equivalently, anisotropy) structure.
Here we also note that Finsler like metrics could be modelled in
 the framework of (pseudo)Riemannian geometry (under well defined
 conditions some Finsler metrics could be solutions of canonical Einstein
 equations) if dynamical reductions from higher dimensions to low
 dimensional ones are considered.

Since the seminal works of Bekenstein \cite{bek}, Bardeen, Carter and
Hawking \cite{bch} and Hawking \cite{haw1}, black holes were shown to have
properties very similar to those of ordinary thermodynamics. One was treated
the surface gravity on the event horizon as the temperature of the black
hole and proved that a quarter of the event horizon area corresponds to the
entropy of black holes. At present time it is widely believed that a black
hole is a thermodynamic system (in spite of the fact that one have been
developed a number of realizations of thermodynamics involving radiation)
and the problem of statistical interpretation of the black hole entropy is
one of the most fascinating subjects of modern investigations in
gravitational and string theories.

In parallel to the 'thermodynamilazation' of black hole physics one have
developed a new approach to the classical thermodynamics based of Riemannian
geometry and its generalizations (a review on this subject is contained in
Ref. \cite{rup}). Here is to be emphasized that geometrical methods have
always played an important role in thermodynamics (see, for instance, a work
by Blaschke \cite{bla} from 1923). Buchdahl used in 1966 a Euclidean metric
in thermodynamics \cite{buch} and then Weinhold considered a sort of
Riemannian metric \cite{wein}. It is considered that the Weinhold's metric
has not physical meanining in the context of purely equilibrium
thermodynamics \cite{rup0,rup} and Ruppeiner introduced a new metric
(related via the temperature $T$ as the conformal factor with the Weinhold's
metric).

The  thermodynamical geometry was generalized in various directions,
for instance, by Janyszek and Mrugala \cite{jm1,jm2,m3} even to
discussions of applications of Finsler geometry in thermodynamic
fluctuation theory and for nonequilibrium thermodynamics \cite{sien}.

Our goal will be to provide a characterization of thermodynamics of
 $2+1$--dimensional black la--holes with elliptical (constant in time)
 horizon obtained in \cite{v3}. From one point of view we shall consider
 the thermodynamic space of such objects (black la--holes in local
 equilibrium with la--spacetime ether) to depend  on parameter of
  anisotropy, the angle $\theta ,$ and on deformation parameter, the
 eccentricity $\varepsilon .$ From another point,
 after we shall integrate the formulas on $\theta ,$
 the thermodynamic  geometry will be considered in a usual
  two--dimensional   Ruppeiner--Mrugala--Janyszek fashion.
 The main result of this work are the
 computation of thermodynamic curvatures and the
 proof that constant in time elliptic black la--holes have critical
 points of vanishing of curvatures (under both approaches to
 two--dimensional thermodynamic geometry) for some values of
 eccentricity, i. e. for under corresponding deformations of
 la--spacetimes.

The paper is organized as follows. In Sec. II, we briefly review
 the geometry la--spacetimes provided with nonlinear connection
 structure and present the  $2+1$--dimensional constant in time
 elliptic black la--hole solution. In Sec. III, we state the
 thermodynamics of nearly equilibrium stationary black la--holes
 and establish the basic thermodynamic law and relations.
 In Sec. IV we develop two approaches to the thermodynamic geometry
 of black la--holes, compute thermodynamic curvatures and
 the equations for critical points of vanishing of curvatures
 for some values of eccentricity. In Sec. V, we draw
 a discussion and conclusions.

\section{Locally Anisotropic Spacetimes and Black Holes}

In this section we outline for further applications
the basic results on $2+1$--dimensional la--spacetimes and
 black la--hole solutions (see Refs. \cite{v3,v4} and the
 geometric background presented in Appendix; for
$(2+1)$--anisotropies all formula are considered for
 dimensions of base spaces $n=3$ and of fibers $m=1).$

\subsection{Nonlinear Connections in $(2+1)$--Dimensional Spacetimes}

A (2+1)--dimensional locally anisotropic spacetime is defined as a generic
anholonomic vector bundle with an one dimensional fiber, parametrized by a
coordinate $y,$ over a two dimensional base space, locally parametrized by
coordinates $x^i,i=1,2.$  We
shall use also the next denotations of coordinates $
u=(x,y)=\{u^\alpha =(x^i,y)\}. $

The local anisotropy (la) is modelled by the coefficients of nonlinear
connection (in brief, N--connection)
$$
N=\{N_i\left( x,y\right) =N_i^b\left( x^j,y\right) \},
$$
were, for simplicity, for N--connection components there are omitted
 the one dimensional fiber indices. Nonlinear connections generalize
the concept of usual linear connections and can be treated as a field
splitting the generic anholonomic spacetime into irreducible horizontal
(base) and vertical (anisotropy) subspaces. The components of N--connection
could be considered as prescribed values (functions)
if some constraints on spacetime dynamics are imposed,
 or as coefficients of a specific  nonlinear gauge field
satisfying corresponding motion (field) equations.
 On generic la--spacetimes one have to apply
'elongated' by N--connections operators instead of usual local coordinate
basis $\partial _\alpha =\partial /\partial u^\alpha $ and $d^\alpha
=du^\alpha ,$ (see formulas (A3) and (A4)):%
\begin{eqnarray}
\delta _\alpha & = &
(\delta _i,\partial _{(y)})=\frac \delta {\partial u^\alpha }%
\eqnum{2.1}
 \\
 & = &
 \left( \frac \delta {\partial x^i}=\frac \partial {\partial x^i}-
N_i\left(
x^j,y\right) \frac \partial {\partial y},
 \partial _{(y)} = \frac \partial {\partial y}\right)
\nonumber
\end{eqnarray}
and their duals%
\begin{eqnarray}
\delta ^\beta & = & \left( d^i,\delta ^{(y)}\right)
  =  \delta u^\beta \eqnum{2.2}
\\ & = &
\left( d^i = dx^i,
 \delta ^{(y)} = \delta y=dy+N_k\left( x^j,y\right) dx^k\right) .
\nonumber
\end{eqnarray}

The locally adapted to the N--connection structure operators of partial
derivatives (2.1) and their duals, differentials, (2.2) form generic
anholonomic frames, or la--bases. With respect to a fixed structure of
la--bases and their tensor products we can construct distinguished, by
N--connection, tensor algebras and various geometric objects (in brief, one
writes d--tensors, d--metrics, d--connections and so on).

A symmetrical la--metric, or d--metric, could be written with respect to a
la--basis (2.2) as%
\begin{eqnarray}
\delta s^2 &= &
g_{\alpha \beta }\left( u^\tau \right)
\delta u^\alpha \delta u^\beta
\eqnum{2.3}\\
& = & g_{ij}\left( x^k,y\right) dx^idx^j+h\left( x^k,y\right)
 \left( \delta
y\right) ^2.
\nonumber
\end{eqnarray}
Such metrics have been used in generalized Finsler and Lagrange geometries
\cite{ma} and for modelling Finsler--Kaluza--Klein (super)gravities on
(super)vector bundles provided with N--connection structures \cite{v1,v2,v3}.

\subsection{Non--rotating black la--holes with ellipsoidal horizon}

Let us consider a la--spacetime provided with local coordinates $x^1=r,\ x^2
= \theta$ for the base subspace and when as the anisotropic direction is
chosen the time like coordinate, $y= t.$ We proved \cite{v4} that a
d--metric of type (2.3),
$$
\delta s^2=g\left( r,\theta \right) dr^2+r^2d\theta ^2+ h(r,\theta )\delta
t^2,\eqno{(2.4)}
$$
where%
$$
\delta t=dt+N_1\left( r,\theta \right) dr+N_2(r,\theta )d\theta ,
$$
satisfies the system of vacuum la--gravitational equations in (2+1)
dimensions if
$$
h(r,\theta )=\frac 1{g\left( r,\theta \right) }= -\frac{p^2}{\left(
1+\varepsilon \cos \left( \theta -\theta _0\right) \right) ^2}+ \frac{r^2}{%
r_0^2}, \eqno{(2.5)}
$$
where $p,\varepsilon ,\theta _0$ and $r_0$ are constants and the
N--connection has the coefficients
\begin{eqnarray}
N_1 & = &H\frac{\partial g}{\partial r}
\left[ 2r\frac{\partial g}{\partial r}%
+\left( \frac{\partial g}{\partial \theta}\right) ^2\right] ^{-1},
 \eqnum{2.6}  \\
N_2 & = & H
\frac{\partial g}{\partial \theta}
\left[ 2r\frac{\partial g}{\partial r}%
+\left( \frac{\partial g}{\partial \theta}\right) ^2\right] ^{-1},
\nonumber
\end{eqnarray}
where%
$$
\frac H2=\frac{\partial ^2g}{\partial \theta ^2}-\frac 1{2g} \left( \frac{%
\partial g}{\partial \theta}\right) ^2-\frac rg\frac{\partial g}{\partial r}%
.
$$

The time--time component $h(r,\theta )$ vanishes if
$$
r_{+ }^2= \frac{p^2r_0^2}{\left( 1+\varepsilon \cos \left( \theta -\theta
_0\right) \right) ^2} \eqno{(2.7)}
$$
which is the square of the parametric equation of a ellipse with parameter $%
p $ and eccentricity $\varepsilon $ and where the angle $\theta _0$ gives
the orientation of axes. If we impose the condition that in the locally
isotropic limit we shall have the usual BTZ solution, we can express the
constants $p$ and $r_0$ via the standard mass and cosmological constant, i.
e. $p^2=m_0$ and $r_0^2=-1/\Lambda .$ The eccentricity $\varepsilon $ and
axes orientation $\theta _0$ are given by the initial conditions of
la--gravitational space polarization.

We are thus led to the result that
 a d--metric (2.4), with coefficients (2.5) and for the N--connection (2.6),
describes a locally anisotropic variant of the Schwarzschild metric and has
an ellipsoidal horizon.

\section{On the Thermodynamics of Elliptical Black La--Holes}

In this paper we will be interested in thermodynamics of black la--holes
 defined by a d--metric (2.4).

The coefficient before $r_0^2$ in (2.7) can be treated as an anisotropic mass%
$$
m\left( \theta ,\varepsilon \right) =\frac{m_0}{2\pi \left( 1+\varepsilon
\cos \left( \theta -\theta _0\right) \right) ^2}=\frac{r_{+}^2}{2\pi r_0^2}%
\eqno{(3.1)}
$$
which depends on coordinate $\theta $  and eccentricity $%
\varepsilon $ and on constants $m_0$ and $\theta _0.$ The coefficient $2\pi
\,$ was introduced in order to have the limit
$$
\lim _{\varepsilon \rightarrow 0}2\int\limits_0^\pi m\left( \theta
,\varepsilon \right) d\theta =m_0.\eqno{(3.2)}
$$
Throughout this paper, the units $c=\hbar =k_B=1$ will be used, but we shall
consider that for an la--renormalized gravitational constant $%
8G_{(gr)}^{(a)}\neq 1,$ see \cite{v4}.

The Hawking temperature $T\left( \theta ,\varepsilon \right) $ of a black
la--hole is anisotropic and is computed by using the anisotropic mass (3.1):%
$$
T\left( \theta ,\varepsilon \right) =\frac{m\left( \theta ,\varepsilon
\right) }{2\pi r_{+}\left( \theta ,\varepsilon \right) }=\frac{r_{+}\left(
\theta ,\varepsilon \right) }{4\pi ^2r_0^2}>0.\eqno{(3.3)}
$$

The two parametric analog of the Bekenstein--Hawking entropy is to be
defined as%
$$
S\left( \theta ,\varepsilon \right) =4\pi r_{+}=\sqrt{32\pi ^3}|r_0|\sqrt{%
m\left( \theta ,\varepsilon \right) }\eqno{(3.4)}
$$

The introduced thermodynamic quantities obey the first law of thermodynamics
(under the supposition that the system is in local equilibrium under the
variation of parameters $\left( \theta ,\varepsilon \right) $)
$$
\Delta m\left( \theta ,\varepsilon \right) =T\left( \theta ,\varepsilon
\right) \Delta S, \eqno{(3.5)}
$$
where the variation of entropy is
$$
\Delta S=4\pi \Delta r_{+}=4\pi \frac{\left| r_0\right| }{\sqrt{m\left(
\theta ,\varepsilon \right) }}\left( \frac{\partial m}{\partial \theta }%
\Delta \theta +\frac{\partial m}{\partial \varepsilon }\Delta \varepsilon
\right) .
$$
According to the formula $C=\left( \partial m/\partial T\right) $ we can
compute the heat capacity%
$$
C=2\pi r_{+}\left( \theta ,\varepsilon \right) =2\pi \left| r_0\right| \sqrt{%
m\left( \theta ,\varepsilon \right) }.
$$
Because of $C>0$ always holds the temeperature is increasing with the mass.

The formulas (3.1)-(3.5) can be integrated on angular variable $\theta $ in
order to obtain some thermodynamic relations for black la--holes with
elliptic horizon depending only on deformation parameter, the eccentricity $%
\varepsilon .$

For a elliptically deformed black la-hole with the outer horizon $r_{+}$
given by formula (3.4) the depending on eccentricity\cite{v4}
Bekenstein--Hawking entropy is computed as
$$
S^{(a)}\left( \varepsilon \right) =\frac{L_{+}}{4G_{(gr)}^{(a)}},
$$
were
$$
L_{+}\left( \varepsilon \right) =4\int\limits_0^{\pi /2}r_{+}\left( \theta
,\varepsilon \right) d\theta
$$
is the length of ellipse's perimeter and $G_{(gr)}^{(a)}$ is the three
dimensional gravitational coupling constant in la--media (the index $\left(
a\right) $ points to la--renormalizations), and has the value
$$
S^{(a)}\left( \varepsilon \right) =\frac{2pr_0}{G_{(gr)}^{(a)}\sqrt{%
1-\varepsilon ^2}}arctg\sqrt{\frac{1-\varepsilon }{1+\varepsilon }}.%
\eqno{(3.6)}
$$
If the eccentricity vanishes, $\varepsilon =0,$ we obtain the locally
isotropic formula with $p$ being the radius of the horizon circumference,
but the constant $G_{(gr)}^{(a)}$ could be la--renormalized.

The total mass of black la--hole of eccentricity $\varepsilon $ is found by
integrating (3.1) on angle $\theta :$%
$$
m\left( \varepsilon \right) =\frac{m_0}{\left( 1-\varepsilon ^2\right) ^{3/2}%
}\eqno{(3.7)}
$$
which satisfies the condition (3.2).

The integrated on angular variable $\theta $ temperature $T\left(
\varepsilon \right) $ is to by defined by using $T\left( \theta ,\varepsilon
\right) $ from (3.3),
$$
T\left( \varepsilon \right) =4\int\limits_0^{\pi /2}T\left( \theta
,\varepsilon \right) d\theta =\frac{2\sqrt{m_0}}{\pi ^2|r_0|\sqrt{%
1-\varepsilon ^2}}arctg\sqrt{\frac{1-\varepsilon }{1+\varepsilon }.}%
\eqno{(3.8)}
$$

Formulas (3.6)-(3.8) describes the thermodynamics of $\varepsilon $%
--deformed black la--holes.

Finally, in this section, we note that a black la--hole with elliptic
horizon is to be considered as a thermodynamic subsystem placed into the
la-ether bath of spacetime. To the la--ether one associates a continuous
la--medium assumed to be in local equilibrium. The black la--hole subsystem
is considered as a subsystem described by thermodynamic variables which are
continuous field on variables $\left( \theta ,\varepsilon \right) ,$ or in
the simplest case when one have integrated on $\theta ,$ on $\varepsilon .$
It will be our first task to establish some parametric thermodynamic
relations between the mass $m\left( \theta ,\varepsilon \right) $
(equivalently, the internal black la--hole energy), temperature $T\left(
\theta ,\varepsilon \right) $ and entropy $S\left( \theta ,\varepsilon
\right) .$

\section{Thermodynamic Metrics and Curvatures of Black La--Holes}

 We emphasize in this paper
 two approaches to the thermodynamic geometry of nearly
 equilibrium black la--holes based on their thermodynamics. The first one
 is to consider  the thermodynamic space as depending locally on two
 parameters $\theta$ and $\varepsilon$ and to compute  the
 corresponding metric and curvature following standard formulas from
 curved bidimensional hypersurface Riemannian geometry. The second
 possibility is to take as basic the Ruppeiner metric in the thermodynamic
 space with coordinates $(M, \varepsilon ),$ in a manner
 proposed in Ref. \cite{cai} with that difference that as the extensive
 coordinate is taken the black la--hole eccentricity $\varepsilon$ (instead
 of the usual angular momentum $J$ for isotropic $(2+1)$--black holes).
 Of course, in this case we shall background our thermodynamic geometric
 constructions starting from the relations (3.6)--(3.8).

\subsection{The thermodynamic parametric geometry}

Let us consider  the thermodynamic parametric geometry of the elliptic
(2+1)--dimensional black la--hole based on its thermodynamics given
 by formulas (3.1)--(3.5).

Rewriting equations (3.5), we have%
$$
\Delta S=\beta \left( \theta ,\varepsilon \right) \Delta m\left( \theta
,\varepsilon \right) ,
$$
where $\beta \left( \theta ,\varepsilon \right) =1/T\left( \theta
,\varepsilon \right) $ is the inverse to temperature (3.3). This case is
quite different from that from \cite{cai,ferrara} where there are
considered, respectively, BTZ and dilaton black holes (by introducing
Ruppeiner and Weinhold thermodynamic metrics). Our thermodynamic space is
defined by a hypersurface given by parametric dependencies of mass and
entropy. Having chosen as basic the relative entropy function,
$$
\varsigma =\frac{S\left( \theta ,\varepsilon \right) }{4\pi \sqrt{m_0}}=%
\frac 1{1+\varepsilon \cos \theta },
$$
in the vicinity of a point $P=(0,0),$ when, for simplicity, $\theta _0=0,$
our hypersurface is given locally by conditions%
$$
\varsigma =\varsigma \left( \theta ,\varepsilon \right)
$$
and
$$
grad_{|P}\varsigma =0.
$$
For the components of bidimensional metric on the hypersurface we have%
\begin{eqnarray}
g_{11} & = & 1+\left( \frac{\partial \varsigma }
{\partial \theta }\right) ^2, \
g_{12}  =  \left( \frac{\partial \varsigma }{\partial \theta }\right)
 \left(\frac{\partial \varsigma }{\partial \varepsilon }\right) ,
\nonumber \\
g_{22} & = & 1+
\left( \frac{\partial \varsigma }{\partial \varsigma }\right) ^2, \nonumber
\end{eqnarray}
The nonvanishing component of curvature tensor in
the vicinity of the point $P=(0,0)$ is
$$
R_{1212}=\frac{\partial ^2\varsigma }{\partial \theta ^2}\frac{\partial
^2\zeta }{\partial \varepsilon ^2}-\left( \frac{\partial ^2\varsigma }{%
\partial \varepsilon \partial \theta }\right) ^2
$$
and the curvature scalar is
$$
R=2R_{1212}.\eqno{(4.1)}
$$

By straightforward calculations we can find the condition of vanishing of the
curvature (4.1) when
$$
\varepsilon _{\pm }=\frac{-1\pm (2-\cos ^2\theta )}{\cos \theta \left(
3-\cos ^2\theta \right) }.\eqno{(4.2)}
$$
So, the parametric space is separated in subregions with elliptic
eccentricities $0<\varepsilon _{\pm }<0$ and $\theta $ satisfying conditions
(4.2).

Ruppeiner suggested that the curvature of thermodynamic space is a measure
of the smallest volume where classical thermodynamic theory based on the
assumption of a uniform environment could conceivably work and that near the
critical point it is expected this volume to be proportional to the scalar
curvature \cite{rup}. There were also proposed geometric equations relating
the thermodynamic curvature via inverse relations to free energy. Our
definition of thermodynamic metric and curvature in parametric spaces
differs from that of Ruppeiner or Weinhold and it is obvious that relations
of type (4.2) (stating the conditions of vanishing of curvature) could be
related with some conditions for stability of thermodynamic space under
variations of eccentricity $\varepsilon $ and anisotropy angle $\theta .$
This interpretation is very similar to that proposed by Janyszek and Mrugala
\cite{jm1} and supports the viewpoint that the first law of thermodynamics
makes a statement about the first derivatives of the entropy, the second law
is for the second derivatives and the curvature is a statement about the
third derivatives. This treatment holds good also for the parametric
thermodynamic spaces for black la--holes.

\subsection{Thermodynamic Metrics and Eccentricity of Black La--Hole}

A variant of thermodynamic geometry of black la--holes could be backgrounded
on integrated on anisotropy angle $\theta $ formulas (3.6)-(3.8). The
Ruppeiner metric of elliptic black la--holes in coordinates $\left(
m,\varepsilon \right) $ is
$$
ds_R^2=-\left( \frac{\partial ^2S}{\partial m^2}\right) _\varepsilon
dm^2-\left( \frac{\partial ^2S}{\partial \varepsilon ^2}\right)
_md\varepsilon ^2.\eqno{(4.3)}
$$
For our further analysis we shall use dimensionless values $\mu =m\left(
\varepsilon \right) /m_0$ and $\zeta =S^{(a)}G_{gr}^{(a)}/2pr_0$ and
consider instead of (4.3) the thremodynamic diagonal metrics $g_{ij}\left(
a^1,a^2\right) =g_{ij}(\mu ,\varepsilon )$ with components
$$
g_{11}=-\frac{\partial ^2\zeta }{\partial \mu ^2}=-\zeta _{,11}
\mbox{\ and \ }
g_{22}=-\frac{\partial ^2\zeta }{\partial \varepsilon ^2}=-\zeta _{,22},
\eqno{(4.4)}
$$
where by comas we have denoted partial derivatives.

The expressions (3.6) and (3.7) are correspondingly rewritten as%
$$
\zeta =\frac 1{\sqrt{1-\varepsilon ^2}}arctg\sqrt{\frac{1-\varepsilon }{%
1+\varepsilon }}
$$
and
$$
\mu =\left( 1-\varepsilon ^2\right) ^{-3/2}.
$$

By straightforward calculations we obtain
\begin{eqnarray}
\zeta _{,11} & = &
-\frac 19\left( 1-\varepsilon ^2\right) ^{5/2}arctg\sqrt{\frac{%
1-\varepsilon }{1+\varepsilon }} \nonumber \\
{ } & + & \frac 1{9\varepsilon }\left( 1-\varepsilon
^2\right) ^3+\frac 1{18\varepsilon ^4}\left( 1-\varepsilon ^2\right) ^4
\nonumber
\end{eqnarray}
and
$$
\zeta _{,22}=\frac{1+2\varepsilon ^2}{\left( 1-\varepsilon ^2\right) ^{5/2}}%
arctg\sqrt{\frac{1-\varepsilon }{1+\varepsilon }}-\frac{3\varepsilon }{%
\left( 1-\varepsilon ^2\right) ^2}.
$$

The thermodynamic curvature of metrics of type (4.4) can be written in terms
of second and third derivatives \cite{jm1} by using third and second order
determinants:%
\begin{eqnarray}
R & = & \frac 12\left|
\begin{array}{ccc}
-\zeta _{,11} & 0 & -\zeta _{,22} \\
-\zeta _{,111} & -\zeta _{,112} & 0 \\
-\zeta _{,112} & 0 & -\zeta _{,222}
\end{array}
\right| \times \left|
\begin{array}{cc}
-\zeta _{,11} & 0 \\
0 & -\zeta _{,22}
\end{array}
\right| ^{-2} \nonumber \\
 {} & = &
-\frac 12\left( \frac 1{\zeta _{,11}}\right) _{,2}\times \left( \frac{\zeta
_{,11}}{\zeta _{,22}}\right) _{,2}. \eqnum{4.5}
\end{eqnarray}
The conditions of vanishing of thermodynamic curvature (4.5) are as follows%
$$
\zeta _{,112}\left( \varepsilon _1\right) =0\eqno{(4.6)}
$$
or%
$$
\left( \frac{\zeta _{,11}}{\zeta _{,22}}\right) _{,2}\left( \varepsilon
_2\right) =0\eqno{(4.7)}
$$
for some values of eccentricity, $\varepsilon =\varepsilon _1$ or $%
\varepsilon =\varepsilon _2,$ satisfying conditions $0<\varepsilon _1<1$ and
$0<\varepsilon _2<1.$ For small deformations of black la--holes, i.e. for
small values of eccentricity, we can approximate $\varepsilon _1\approx 1/%
\sqrt{5.5}$ and $\varepsilon _2\approx 1/(18\lambda ),$ where $\lambda $ is
a constant for which $\zeta _{,11}=\lambda \zeta _{,22}$ and the condition $%
0<\varepsilon _2<1$ is satisfied. We omit general formulas for curvature
(4.5) and conditions (4.6) and (4.7), when the critical points $\varepsilon
_1$ and/or $\varepsilon _2$ must be defined from nonlinear equations
containing $arctg\sqrt{\frac{1-\varepsilon }{1+\varepsilon }}$ and powers of
$\left( 1-\varepsilon ^2\right) $ and $\varepsilon .$

\section{Discussion and Conclusions}

In closing, we would like to discuss the meaning of geometric
 thermodynamics following from black locally anisotropic (la) holes
 (in brief we shall use la--holes, la--spacetime and so on).

(1) {\it Nonequilibrium thermodynamics of black la--holes in
 la--spacetimes}. In this paper and in paper I \cite{v4} we
 concluded that the thermodynamics of generic la--spacetimes has a generic
 nonequilibrium character and could be developed in a geometric
 fashion following the approach proposed by S. Sieniutycz, P. Salamon
 and R. S. Berry \cite{sien,sal}. This could form a new branch
 of black hole thermodynamics which should be based on previous
 studies in nonequilibrium thermodynamics and kinetics, theirs
 stochastic \cite {v6} and statistical background, and on thermodynamical
 hydrodynamics. All constructions will be performed on generic
 anholonomic manifolds by modelling la--spacetimes
 on vector bundles provided with nonlinear connection structure.
  This direction could be motivated after it was found that
 generalized and standard Finsler like metrics could be considered
 in the framework of Einstein gravity under corresponding parametrizations
 of metrics and reductions from higher dimensions to low dimensional
 ones.

(2) {\it Black la--holes thermodynamics in vicinity of equilibrium
 points}. The usual thermodynamical approach in the Bekenstein--Hawking
 manner is valid for black la--holes for a subclass of such physical
 systems when the hypothesis of local equilibrium is physically motivated
 and corresponding renormalizations, by la--spacetime parameters,
  of thermodynamical values are defined.

(3) {\it The geometric thermodynamics of black la--holes with constant
 in time elliptic horizon} was formulated following two approaches:
  for a parametric
 thermodynamic space depending on anisotropy angle $\theta$ and
 eccentricity $\varepsilon$ and in a standard Ruppeiner--Mrugala--Janyszek
 fashion, after integration on anisotropy $\theta$ but maintaining
 la--spacetime deformations on $\varepsilon .$

(4) {\it The thermodynamic curvatures of black la--holes} were shown to
 have critical values of eccentricity when the scalar curvature vanishes.
 Such type of thermodynamical systems are rather unusual and a corresponding
 statistical model is not that for  ordinary systems composed by
 classical or quantum like gases.

(5) {\it Thermodynamic systems with constraints} requires a new geometric
 structure in addition to the thermodynamical metrics which is
 that of nonlinear  connection. We note this object
must be introduced both in
 spacetime geometry and in thermodynamic geometry if generic anisotropies
 and constrained field and/or thermodynamic behavior are analyzed.

\acknowledgments
The author is grateful to Professors R. Mrugala and M. Michalski for
 helpful discussions, support and hospitality during his 
 at Torun, Poland.

\appendix

\section{ Anholonomic Bundles and Locally Anisotropic Spaces}

We outline the geometry of anholonomic bundles and of the N--connection
structure for locally anisotropic spaces \cite{haw} and \cite{ma,v1,v2}.

\subsection{Anholonomic manifolds and gravity}

In this section all manifolds $E_{d_{(E)}}$ are assumed to be smooth (i.e. $%
C^\infty ),$ of finite integer dimension $d_{(E)}\geq 3,4,...,$ Hausdorff,
paracompact and connected; all maps are smooth. We denote the local
coordinates on $E_{d_{(E)}}$ by variables $u^\alpha ,$ where Greek indices
takes values $\alpha ,\beta ,...=1,2,3,4,...$ and could be both type
coordinate or abstract (Penrose's) ones. A spacetime is modelled by a
manifold $E_{d_{(E)}}$ provided with corresponding geometric structures
(symmetric metric $g_{\alpha \beta };$ linear, in general nonsymmetric,
connection $\Gamma _{~\beta \gamma }^\alpha $ defining the covariant
derivation $\nabla _\alpha ;$ nonmetricity $Q_{\alpha \beta \gamma }=\nabla
_\alpha g_{\beta \gamma }$ which in this work is considered to be vanishing,
i.e.$Q_{\alpha\beta \gamma }\equiv 0)$ . If it would be necessary to
emphasize that some indices are abstract marks, we shall underline them,
i.e. we shall write $\underline{\alpha },\underline{\beta },....$

Frame basis vectors on $E_{d_{(E)}},$ numbered by a index $\underline{\alpha
}$ are denoted by $e^{\underline{\alpha }}$ with components $e_\alpha ^{%
\underline{\alpha }}=g_{\alpha \beta }e^{\underline{\alpha }\beta },$ i.e. $%
e^{\underline{\alpha }}=\{e_\alpha ^{\underline{\alpha }}=g_{\alpha \beta
}e^{\underline{\alpha }\beta }\},$ and they are subjected to relations%
$$
e_{\underline{\alpha }}^\alpha e_{\underline{\beta }\alpha }=\eta _{%
\underline{\alpha },\underline{\beta }},
\mbox{ \ and \ }
e_\alpha ^{\underline{\alpha }}e_\beta ^{\underline{\beta }}\eta _{%
\underline{\alpha },\underline{\beta }}=g_{\alpha \beta }\eqno{(A1)}
$$
where the Einstein summations rule is accepted, $\eta _{\underline{\alpha },%
\underline{\beta }}$ is a given constant symmetric matrix of signature $%
\left( -,+....+\right) $ (the sign minus is used in this work for the time
coordinate of spacetime). Operations with underlined and non--underlined
indices are correspondingly performed by using the matrix $\eta _{\underline{%
\alpha },\underline{\beta }},$ its inverse $\eta ^{\underline{\alpha }%
\underline{\beta }},$ and the metric $g_{\alpha \beta }$ and its inverse $%
g^{\alpha \beta }.$ A frame basis structure on $E_{d_{(E)}}$ is
characterized by its anholonomy coefficients $w_{~\beta \gamma }^\alpha $
defined from relations%
$$
e_\alpha e_\beta -e_\beta e_\alpha =w_{~\alpha \beta }^\gamma e_\gamma .%
\eqno{(A2)}
$$

With respect to a fixed basis $e_\alpha $ and its dual $e^\beta $ on $%
E_{d_{(E)}}$ we can decompose tensors and define their components, for
instance,%
$$
T=T_{~\alpha \beta }^\gamma ~e_\gamma \otimes e^\alpha \otimes e^\beta
$$
where by $\otimes $ it is denoted the tensor product.

We can also consider local linear transforms of frames,\
$
e_{\underline{\alpha }}=a_{\underline{\alpha }}^{\underline{\alpha ^{\prime }%
}}e_{\underline{\alpha }^{\prime }}, $\
parametrized by nondegenerated linear matrices $a_{\underline{\alpha }}^{%
\underline{\alpha ^{\prime }}}$ and define the corresponding linear frame
bundle on $E_{d_{(E)}},$ or introduce local affine transforms of frames,
$e_{\underline{\alpha }} =
a_{\underline{\alpha }}^{\underline{\alpha ^{\prime }%
}}e_{\underline{\alpha }^{\prime }}+q_{\underline{\alpha }},
$\
with additional affine shifts given by $q_{\underline{\alpha }}$ and define
the corresponding affine frame bundle on $E_{d_{(E)}}.$

Now we discuss the difference between the anholonomic and holonomic
manifolds. A spacetime $E_{d_{(E)}}$ is{\bf \ holonomic (locally integrable)}
if it admits a frame structure for which the anholonomy coefficients from
(A2) vanishes, i.e. $w_{~\alpha \beta }^\gamma =0.$ In this case we can
introduce local coordinate bases,
$$
\partial _\alpha =\partial /\partial u^\alpha \eqno{(A3)}
$$
and their duals
$$
d^\alpha =du^\alpha \eqno{(A4)}
$$
and consider decompositions of geometrical objects with respect to such
frames. The general relativity theory was formally defined for holonomic
pseudo-Riemannian manifold. For various purposes on holonomic spacetimes it
is convenient to use anholonomic frames $e_{\underline{\alpha }}$ satisfying
conditions (A2), but we emphasize that for such spaces we can always define
linear transforms of frames to a coordinate basis of type (A3), $e_{%
\underline{\alpha }}=a_{\underline{\alpha }}^{\underline{\alpha ^{\prime }}%
}\partial _{\underline{\alpha }^{\prime }}.$ By applying frames and theirs
transforms on holonomic pseudo-Riemannian spaces there were developed the
so-called tetradic and spinor gravity and extensions to linear, affine and
de Sitter gauge group gravity models.

A spacetime $E_{d_{(E)}}$ is{\bf \ anholonomic (locally non-integrable)} if
it does not admit a frame structure for which the anholonomy coefficients
from (A1) vanishes, i.e. $w_{~\alpha \beta }^\gamma \neq 0.$ In this case
the anholonomy becomes a generic spacetime characteristics. It induces
nonvanishing additional terms into the torsion, $T\left( \delta _\gamma
,\delta _\beta \right) =T_{~\beta \gamma }^\alpha \delta _\alpha ,$ and
curvature, $R\left( \delta _\tau ,\delta _\gamma \right) \delta _\beta
=R_{\beta ~\gamma \tau }^{~\alpha }\delta _\alpha ,$ tensors of a linear
connection $\Gamma _{~\beta \gamma }^\alpha ,$ with coefficients defined
respectively as
$$
T_{~\beta \gamma }^\alpha =\Gamma _{~\beta \gamma }^\alpha -\Gamma _{~\gamma
\beta }^\alpha +w_{~\beta \gamma }^\alpha \eqno{(A5)}
$$
and
\begin{eqnarray}
R_{\beta ~\gamma \tau }^{~\alpha } & = & \delta _\tau
\Gamma _{~\beta \gamma
}^\alpha -\delta _\gamma \Gamma _{~\beta \delta }^\alpha +
\eqnum{A6}\\
& & \Gamma _{~\beta \gamma }^\varphi \Gamma _{~\varphi \tau }^\alpha
-\Gamma
_{~\beta \tau }^\varphi \Gamma _{~\varphi \gamma }^\alpha +
\Gamma _{~\beta
\varphi }^\alpha w_{~\gamma \tau }^\varphi . \nonumber
\end{eqnarray}

The Ricci tensor is defined as
$$
R_{\beta \gamma }=R_{\beta ~\gamma \alpha }^{~\alpha }\eqno{(A7)}
$$
and the scalar curvature is
$$
R=g^{\beta \gamma }R_{\beta \gamma }\eqno{(A8)}
$$

The Einstein equations on a anholonomic spacetime are introduced in a
standard manner:%
$$
R_{\beta \gamma }-\frac 12g_{\beta \gamma }R=k\Upsilon _{\beta \gamma },%
\eqno{(A9)}
$$
where the energy--momentum d--tensor $\Upsilon _{\beta \gamma }$ includes
the cosmological constant terms and possible contributions of torsion (A5)
and matter and $k$ is the coupling constant. For symmetric linear connection
the torsion field can be considered as induced by the anholonomy (or
equivalently, by imposed constraints). For dynamical torsions there are
necessary additional field equations, see, for instance, the case of gauge
like theories \cite{vg}.

The usual locally isotropic Einstein gravity is obtained under suppositions
that we restrict our considerations only for frame fields (for a four
dimensional spacetime having 16 components) which are locally isotropic and
are locally linearly equivalent to a coordinate vector basis and could
generate via relations (A1) the pseudo--Riemannian metric (having 10
components for a 4 dimensional spacetime).

A subclass of anholonomic spacetimes consists from those with local
anisotropy. It is a matter of further theoretical and experimental
investigations in order to establish if the present day experimental data on
anisotropic structure of Universe is a consequence of matter and quantum
fluctuation induced anisotropies and for some scales being a consequence of
anholonomy of observator's frame, both cases being considered for a locally
isotropic spacetime background, or the spacetime anisotropy is a generic
property following, for instance, from string theory, and from a more
general self--consistent gravitational theory when both the left (geometric)
and right (matter energy--momentum tensor) parts of Einstein equations
depends on anisotropic parameters.

\subsection{The local anisotropy and nonlinear connection}

In this subsection we briefly outline the geometry of locally anisotropic
spaces.

Roughly, a local anisotropy is introduced by calling some space\-time
di\-rec\-ti\-ons (co\-or\-di\-na\-tes) to be an\-isot\-rop\-ic. In this case
the spacetime dimension is split locally into two components, $n$ for
isotropic coordinates and $m$ for anisotropic coordinates$,$ when $%
d_{(E)}=n+m$ with $n\geq 2$ and $m\geq 1.$ We shall use local coordinates $%
u^\alpha =(x^i,y^a),$ where Greek indices $\alpha ,\beta ,...$ take values $%
1,2,...,n+m$ and Latin indices $i$ and $a$ are correspondingly $n$ and $m$
dimensional, i.e. $i,j,k...=1,2,...,n$ and $a,b,c,...=1,2,...,m.$

There is necessary a correct geometric definition of decomposition of
spacetime into isotropic and anisotropic components. For modelling of a
locally anisotropic spacetime, in brief a la--space, we choose a vector
bundle, ${\cal E}=(E_{n+m},p,M_n,F_m,Gr)$ provided with {\bf nonlinear
connection (in brief N--connection)} structure $N=\{N_j^a\left( u^\alpha
\right) \},$ where $N_j^a\left( u^\alpha \right) $ are its coefficients. We
use denotations: $E_{n+m}$ for the $(n+m)$--dimensional total space of the
vector bundle; $M_n$ for the $n$--dimensional base manifold; $F_m$ for the
typical fiber being a $m$--dimensional real vector space; $Gr$ is the group
of automorphisms of $F_m$ and $p$ is a surjective mapping. For simplicity we
shall consider only local constructions on vector bundles.

The N--connection is a new geometric object which generalize that of linear
connection. This concept came from Finsler geometry (see the Cartan's
monograph \cite{cartan}), the global formulation of it is due to W. Barthel
\cite{barthel}, and it is studied in details in Miron and Anastasiei works
\cite{ma}. We have extended the geometric constructions for spinor bundles
and superbundles with further applications in locally anisotropic field
theory and strings .

The rigorous mathematical definition of N--connection is based on the
formalism of horizontal and vertical subbundles and on exact sequences of
vector bundles. Here, for simplicity, we define a N--connection as a
distribution which for every point $u=(x,y)\in {\cal E}$ defines a local
decomposition of the tangent space of our vector bundle, $T_uE,$ into
horizontal, $H_uE,$ and vertical (anisotropy), $V_uE,$ directions, i.e.%
$$
T_uE=H_uE\oplus V_uE.
$$

If a N--connection with coefficients $N_j^a\left( u^\alpha \right) $ is
introduced on the vector bundle ${\cal E}$ the modelled spacetime posses a
generic local anisotropy and in this case we can not even apply in a usual
manner the partial derivatives and their duals, differentials. Instead of
coordinate bases (A3) and (A4) we must consider some bases adapted to the
N--connection structure:
\begin{eqnarray}
\delta _\alpha & = &
(\delta _i,\partial _a ) = \frac \delta {\partial u^\alpha }%
\eqnum{A10}
 \\
 & = &
 \left( \delta _i = \frac \delta {\partial x^i} =
\frac \partial {\partial x^i} -
N_i^b \left( x^j,y\right) \frac \partial {\partial y^b},
 \partial _a  = \frac \partial {\partial y^a}\right)
\nonumber
\end{eqnarray}
and
\begin{eqnarray}
\delta ^\beta & = & \left( d^i,\delta ^a \right)
  =  \delta u^\beta \eqnum{A11}
\\ & = &
\left( d^i = dx^i,
 \delta ^a = \delta y^a =dy^a +N_k^a \left( x^j,y^b \right) dx^k\right) .
\nonumber
\end{eqnarray}

A nonlinear connection (N--connection) is characterized by its curvature%
$$
\Omega _{ij}^a=\frac{\partial N_i^a}{\partial x^j}-\frac{\partial N_j^a}{%
\partial x^i}+N_i^b\frac{\partial N_j^a}{\partial y^b}-N_j^b\frac{\partial
N_i^a}{\partial y^b}.\eqno{(A12)}
$$

The elongation (by N--connection) of partial derivatives, from (A10), called
the adapted to the N--connection partial derivatives, or the locally adapted
basis (la--basis) $\delta _\beta $, reflects the fact that the spacetime is
locally anisotropic ${\cal E}$ and generically anholonomic because there are
satisfied anholonomy relations (A2),
$$
\delta _\alpha \delta _\beta -\delta _\beta \delta _\alpha =w_{~\alpha \beta
}^\gamma \delta _\gamma ,
$$
where anolonomy coefficients are as follows
\begin{eqnarray}
w_{~ij}^k & = & 0,w_{~aj}^k=0,w_{~ia}^k=0,w_{~ab}^k=0,w_{~ab}^c=0,
\nonumber\\
w_{~ij}^a & = &
-\Omega _{ij}^a,w_{~aj}^b=-\partial _aN_i^b,w_{~ia}^b=\partial _aN_i^b.
\nonumber
\end{eqnarray}

On a la--space the geometrical objects have a distinguished (by
N--connection), into horizontal and vertical components, character. They are
briefly called d--tensors, d--metrics and/or d--connections. Their
components are defined with respect to a la--basis of type (A10), it dual
(A11), or their tensor products (d--linear or d--affine transforms of such
frames could also be considered). For instance a covariant and contravariant
d--tensor $Z,$ is expressed as%
\begin{eqnarray}
Z &= & Z_{~\beta }^\alpha \delta _\alpha \otimes \delta ^\beta  \nonumber \\
 {} &= & Z_{~j}^i\delta _i\otimes d^j+Z_{~a}^i\delta _i\otimes \delta
^a+Z_{~j}^b\partial _b\otimes d^j+Z_{~a}^b\partial _b\otimes \delta ^a.
\nonumber
\end{eqnarray}

A symmetric d--metric on la--space ${\cal E}$ is written as
\begin{eqnarray}
\delta s^2 & = & g_{\alpha \beta }\left( u\right)
  \delta ^\alpha \otimes \delta^\beta \eqnum{A13} \\
{} & = & g_{ij}(x,y)dx^idx^j+h_{ab}(x,y)\delta y^a\delta y^b. \nonumber
\end{eqnarray}

A linear d--connection $D$ on la--space ${\cal E,}$
$$
D_{\delta _\gamma }\delta _\beta =\Gamma _{~\beta \gamma }^\alpha \left(
x^k,y\right) \delta _\alpha ,
$$
is parametrized by non--trivial h--v--components,
$$
\Gamma _{~\beta \gamma }^\alpha =\left(
L_{~jk}^i,L_{~bk}^a,C_{~jc}^i,C_{~bc}^a\right) \eqno{(A14)}
$$
Some d--connection and d--metric structures are compatible if there are
satisfied the conditions%
$$
D_\alpha g_{\beta \gamma }=0.
$$
For instance, a canonical compatible d--connection
$$
^c\Gamma _{~\beta \gamma }^\alpha =\left(
^cL_{~jk}^i,^cL_{~bk}^a,^cC_{~jc}^i,^cC_{~bc}^a\right)
$$
is defined by the coefficients of d--metric (A13), $g_{ij}\left( x,y\right) $
and $h_{ab}\left( x,y\right) ,$ and by the coefficients of N--connection,%
\begin{eqnarray}
^cL_{~jk}^i & = & \frac 12g^{in}\left( \delta _kg_{nj}+\delta _jg_{nk}-\delta
_ng_{jk}\right) , \nonumber \\
^cL_{~bk}^a & = & \partial _bN_k^a+\frac 12h^{ac}\left( \delta
_kh_{bc}-h_{dc}\partial _bN_i^d-h_{db}\partial _cN_i^d\right) ,
\nonumber \\
^cC_{~jc}^i & = & \frac 12g^{ik}\partial _cg_{jk}, \nonumber \\
^cC_{~bc}^a & = & \frac 12h^{ad}\left( \partial _ch_{db}+\partial
_bh_{dc}-\partial _dh_{bc}\right)  \nonumber
\end{eqnarray}
This d--connection generalizes for la--spaces the well known Cristoffel
symbols.

For a d--connection (A14) we can compute the components of, in our case
d--torsion, (A5):
\begin{eqnarray}
T_{.jk}^i & = & T_{jk}^i=L_{jk}^i-L_{kj}^i,\quad
T_{ja}^i=C_{.ja}^i,T_{aj}^i=-C_{ja}^i, \nonumber \\
T_{.ja}^i & = & 0,\quad T_{.bc}^a=S_{.bc}^a=C_{bc}^a-C_{cb}^a,
\nonumber \\
T_{.ij}^a & = &
-\Omega _{ij}^a,\quad T_{.bi}^a= \partial _b  N_i^a
-L_{.bj}^a,\quad T_{.ib}^a=-T_{.bi}^a. \nonumber
\end{eqnarray}

In a similar manner, putting non--vanishing coefficients (A14) into the
formula for curvature (A6), we can compute the non--trivial components of
d--curvature
\begin{eqnarray}
R_{h.jk}^{.i} & = & \delta _kL_{.hj}^i-\delta_jL_{.hk}^i \nonumber \\
 & & +  L_{.hj}^mL_{mk}^i-L_{.hk}^mL_{mj}^i-C_{.ha}^i\Omega _{.jk}^a,
\nonumber \\
R_{b.jk}^{.a} & = & \delta _kL_{.bj}^a-\delta_jL_{.bk}^a \nonumber \\
 & & +  L_{.bj}^cL_{.ck}^a-L_{.bk}^cL_{.cj}^a-C_{.bc}^a\Omega _{.jk}^c,
\nonumber \\
P_{j.ka}^{.i} & = & \partial _kL_{.jk}^i +C_{.jb}^iT_{.ka}^b \nonumber \\
 & & -  ( \partial _kC_{.ja}^i+L_{.lk}^iC_{.ja}^l -
L_{.jk}^lC_{.la}^i-L_{.ak}^cC_{.jc}^i ), \nonumber \\
P_{b.ka}^{.c} & = & \partial _aL_{.bk}^c +C_{.bd}^cT_{.ka}^d \nonumber \\
 & & - ( \partial _kC_{.ba}^c+L_{.dk}^{c\,}C_{.ba}^d
- L_{.bk}^dC_{.da}^c-L_{.ak}^dC_{.bd}^c ) \nonumber \\
S_{j.bc}^{.i} & = & \partial _cC_{.jb}^i-\partial _bC_{.jc}^i
 +  C_{.jb}^hC_{.hc}^i-C_{.jc}^hC_{hb}^i, \nonumber \\
S_{b.cd}^{.a} & = &\partial _dC_{.bc}^a-\partial
_cC_{.bd}^a+C_{.bc}^eC_{.ed}^a-C_{.bd}^eC_{.ec}^a. \nonumber
\end{eqnarray}

The components of the Ricci tensor tensor (A7)
$$
R_{\alpha \beta }=R_{\alpha .\beta \tau }^{.\tau }
$$
with respect to locally adapted frames (A9) and (A10) (in our case,
d--tensor) are as follows:%
\begin{eqnarray}
R_{ij} & = & R_{i.jk}^{.k},\quad
 R_{ia}=-^2P_{ia}=-P_{i.ka}^{.k},\eqnum{A15} \\
R_{ai} &= & ^1P_{ai}=P_{a.ib}^{.b},\quad R_{ab}=S_{a.bc}^{.c}. \nonumber
\end{eqnarray}
We point out that because, in general, $^1P_{ai}\neq ~^2P_{ia}$ the Ricci
d-tensor is non symmetric.

Having defined a d-metric of type (A13) in ${\cal E}$ we can introduce the
scalar curvature (A8) of a d-connection $D,$%
$$
{\overleftarrow{R}}=G^{\alpha \beta }R_{\alpha \beta }=\widehat{R}+S,%
\eqno(A16)
$$
where $\widehat{R}=g^{ij}R_{ij}$ and $S=h^{ab}S_{ab}.$

Now, by introducing the values (A15) and (A16) into anholonomic gravity
field equations (A9) we can write down the system of Einstein equations for
la--gravity with prescribed N--connection structure \cite{ma}:%
\begin{eqnarray}
R_{ij}-\frac 12\left( \widehat{R}+S\right) g_{ij} & = &
k\Upsilon _{ij}, \eqnum{A17} \\
S_{ab}-\frac 12\left( \widehat{R}+S\right) h_{ab} & = & k\Upsilon _{ab},
\nonumber \\
^1P_{ai} & = & k\Upsilon _{ai}, \nonumber \\
^2P_{ia} & = & -k\Upsilon _{ia}, \nonumber
\end{eqnarray}
where $\Upsilon _{ij},\Upsilon _{ab},\Upsilon _{ai}$ and $\Upsilon _{ia}$
are the components of the energy--momentum d--tensor field.

There are variants of la--gravitational field equations derived in the
low--energy limits of the theory of locally anisotropic (super)strings \cite
{v2} or in the framework of gauge like la--gravity \cite{vg,v3} when the
N--connection and torsions are dynamical fields.

\end{document}